\documentclass[%
 reprint,
 amsmath,amssymb,
 aps,
]{revtex4-1}

\usepackage[breaklinks=true,colorlinks=true,linkcolor=blue,urlcolor=blue,citecolor=blue]{hyperref}
\usepackage{graphicx}
\usepackage{dcolumn}
\usepackage{bm}
\usepackage[utf8x]{inputenc}

\begin{document}

\preprint{APS/123-QED}

\title{Enhancement of thermoelectric performance in Graphene/BN heterostructures}

\author{Van-Truong Tran}
\thanks{Email: van-truong.tran@u-psud.fr}

\author{Jérôme Saint-Martin }%

\author{Philippe Dollfus}
\thanks{Email: philippe.dollfus@u-psud.fr}
\affiliation{IEF, Université Paris-sud, CNRS, UMR 8622, Bât 220, 91405 Orsay, France.}%

\date{\today}

\begin{abstract}
The thermoelectric properties of in plane heterostructures made of Graphene and hexagonal Boron Nitride (BN) have been investigated by means of atomistic simulation. The heterostructures consist in armchair graphene nanoribbons to the sides of which BN flakes are periodically attached. This arrangement generates a strong mismatch of phonon modes between the different sections of the ribbons, which leads to a very small phonon conductance, while the electron transmission is weakly affected. In combination with the large Seebeck coefficient resulting from the BN-induced bandgap opening or broadening, it is shown that large thermoelectric figure of merit $ZT>0.8$ can be reached in perfect structures at relatively low Fermi energy, depending on the graphene nanoribbon width. The high value $ZT = 1.48$ may even be achieved by introducing appropriately vacancies in the channel, as a consequence of further degradation of the phonon conductance.

\end{abstract}

\maketitle
%

\section{\label{sec:level1}Introduction}

Due to its exceptional electronic, optical, thermal and mechanical properties, graphene should not only replace conventional materials in some existing applications, but, above all the combination of all these unique properties is expected to inspire fully new applications.\cite{Novoselov2012a,C4NR01600A}  However, one major drawback of graphene for electronic applications, at least for its use as a switching device, is its gapless character\cite{CastroNeto2009} that is responsible, e.g., for the low on/off current ratio in graphene transistors.\cite{doi:10.1021/nl102824h,6463444} It is also an obstacle for thermoelectric applications because it makes it difficult to separate the opposite contributions of electron and hole states to the Seebeck coefficient, which is smaller than 100 mV/K in pristine graphene.\cite{Zuev2009} Additionally, the thermal conductivity in graphene is very high, even higher than 4000 W/mK for a single layer,\cite{Balandin2008} which is a strong limitation to achieve high thermoelectric figure of merit $ZT$. Indeed, for small ballistic conductors this quantity is conveniently defined in terms of electronic and thermal conductance as

\begin{equation}
ZT = \frac{P}{K}T
\label{eq_one}
\end{equation}

where $P = {G_e}\,{S^2}$ is the power factor, $K = {K_e} + {K_p}$ is the total thermal conductance, $T$ is the absolute temperature, while ${G_e},{\rm{ }}S,{\rm{ }}{K_e}$ and ${K_p}$ are the electrical conductance, the Seebeck coefficient, the electron thermal conductance and the phonon thermal conductance, respectively. 

\begin{figure}[hbtp]
\centering
\includegraphics[width=9cm, height=3.5cm]{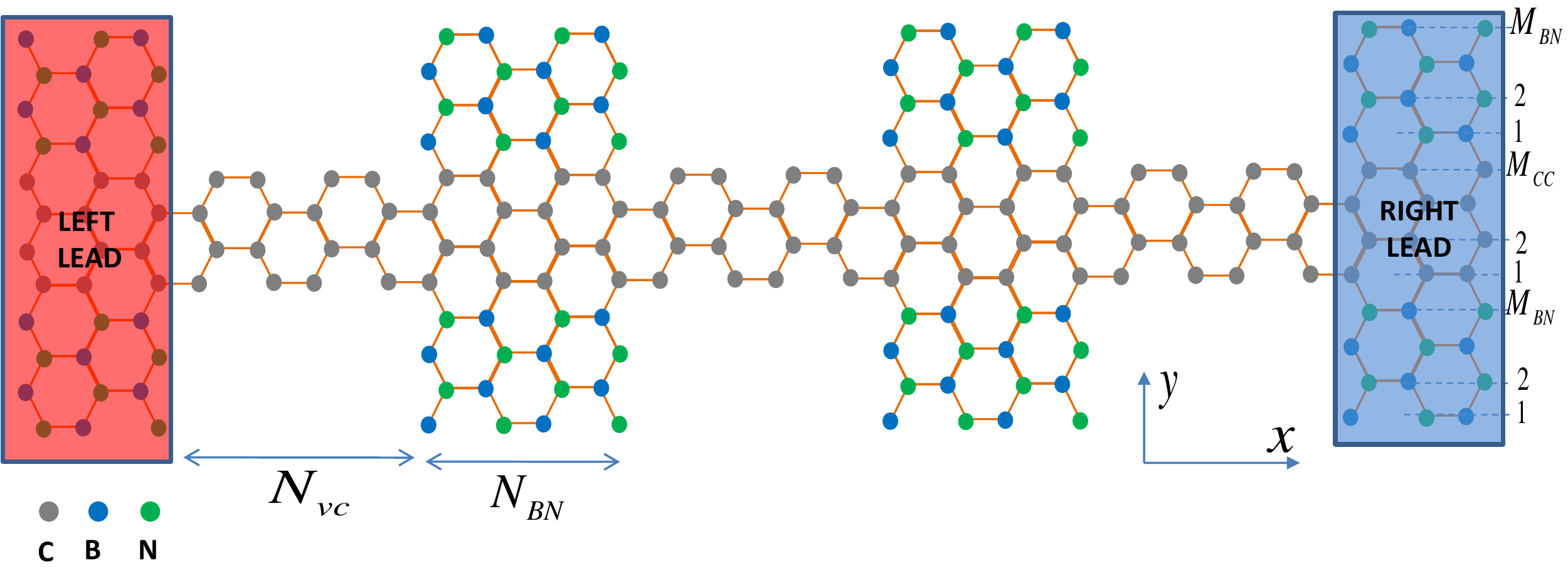}
\caption{Schematic view of graphene/BN heterostructure with sub-sections made of BN flakes attached to the main AGNR channel. The two leads are made of BN/G/BN ribbons as the hybrid parts of the active device.}
\label{Fig_one}
\end{figure}

Actually, the thermoelectric figure of merit in graphene is limited to $ZT < 0.01$, leading to poor thermoelectric efficiency. However, in spite of these intrinsic obstacles, many suggestions of bandgap nanostructuring have been taken up, at least at the theoretical level, to enhance the thermoelectric performance of graphene and to achieve $ZT$ values higher than 1.\cite{0953-8984-27-13-133204} Indeed, since the pioneering works of Hicks and Dresselhaus, nanostructuring materials in low-dimensional systems is strongly expected to provide higher $ZT$ and thermoelectric efficiency than bulk materials,\cite{PhysRevB.47.12727} thanks to (i) the enhancement of Seebeck coefficient resulting from size quantization and (ii) the reduction of thermal conductance due to interface effects. It has been confirmed experimentally in many systems.\cite{ADMA:ADMA201000839} In the case of graphene, the primary expectation of nanostructuring is the bandgap opening to enhance the Seebeck coefficient. The reduction of thermal conductance is strongly expected too, if it is not paid by a stronger reduction of electronic conductance, since all parameters entering the expression of $ZT$ are mutually coupled and difficult to control independently.

As the simplest form of graphene nanostructuring, graphene nanoribbons (GNRs) were first investigated. With finite-bandgap armchair-edges GNRs (AGNRs), significant improvement of thermoelectric properties has been predicted compared to 2D graphene.\cite{Ouyang2009,Mazzamuto2011b} However, in the best case $ZT$ does not exceed 0.35 for the narrowest AGNR with only $M_{CC} = 3$ dimer lines in the width. To enhance further $ZT$, it is necessary to design more sophisticated ribbons likely to degrade strongly the phonon contribution to the thermal conductance while retaining high electron conduction properties. Several works have suggested to use multi-junction GNRs with alternate sections of different width, different chirality, or different chirality and width.\cite{Mazzamuto2011b,Li2014b,Xie2012} Values of $ZT$ close to one have been calculated in such structures. Promising results were also obtained in kinked (or chevron-type) GNRs\cite{:/content/aip/journal/apl/102/14/10.1063/1.4800777} or even by combining chevron-type geometry with isotope cluster engineering.\cite{Sevincli2013} In zigzag-edge GNRs of several micron length, edge disorder and extended line defects have been shown to enhance $ZT$ to values greater than 2 if electron-phonon scattering may remain negligible,\cite{Sevincli2010,Karamitaheri2012} It has been also predicted that $ZT$ can be improved by using graphene nanomesh (GNM), i.e. a graphene layer in which a periodic array of nanopores is perforated\cite{PhysRevB.86.041406} or by designing vertical junctions in multilayer graphene.\cite{Nguyen2014b} 

Recently, a novel form of hybrid monolayer structures consisting of a mixture of graphene and hexagonal-Boron Nitride nanodomains has been proposed and synthesized successfully,\cite{Ci2010d,Vajtai2013,doi:10.1021/nl4021123} thus opening new strategies for bandgap engineering and designing devices. Some theoretical works have demonstrated that Graphene/BN heterostructures made of adjacent armchair ribbons are always semiconducting,\cite{Ding2009a,Seol2011d,Fan2011c,Tang2012a,Jung2012c,Li2013a} while zigzag hybrid ribbons may be either semimetallic or semiconducting, depending on the type of bonding at graphene/BN interfaces,\cite{Fan2011c,Li2013a,Tran2015} which is promising for transistor application with efficient on/off current switching.\cite{Fiori2012a,Tran2014a} Some works also considered the thermal transport in these types of heterostructures\cite{Jiang2011e,Yokomizo2013,Zhu2014} and have shown that they offer the possibility to tune and reduce strongly the phonon thermal conductance. Hence, thanks to the large bandgap opened in Graphene/BN hybrid structures that should lead to good power factor, it is expected that high $ZT$ can be achieved. However, according to our best knowledge only two works have been reported so far on the thermoelectric properties of monolayer graphene/BN structures, by Yang et al.\cite{Yang2012b}, and more recently by Vishkayi et al.\cite{Vishkayi2015}, respectively. The latter focused on hybrid zigzag ribbons with a graphene/BN junction and two leads made of square lattices. Though an enhancement of $ZT$ was observed compared to pure zigzag graphene its maximum value was still limited. In the former work,\cite{Yang2012b} the authors reported a significant improvement of $ZT$ up to 0.7 for some configurations of superlattice armchair ribbons consisting of alternating graphene and BN sections. However, the insulating effect of BN sections induces a high electron scattering, which shifts the good $ZT$ values to high chemical potentials of about 2 eV, i.e. a range of energy difficult to exploit in practice. 

In this article, by means of atomistic calculation we investigate the thermoelectric performance of GNRs with parallel graphene/BN interfaces appropriately distributed. We demonstrate that by graphene/BN interface engineering it is possible to enhance the phonon scattering in armchair GNRs and thus to reduce strongly the phonon thermal conductance while the electronic conductance is weakly affected. It leads to high values of $ZT$ at rather small chemical potential. Additionally, we show that the figure of merit may be enhanced up to 1.48 at room temperature by introducing vacancies in the graphene region. 

The paper is organized as follows. The model and methodology are presented in Sec. \ref{sec:level2}, while the results are discussed in Sec. \ref{sec:level3}, where we emphasize the role of interface phonon scattering (sub-Sec. \ref{sec:level31}) and that of vacancies (sub-Sec. \ref{sec:level32}). The conclusion is in Sec. \ref{sec:level4}.

\section{\label{sec:level2}Studied device and methodolgies }%
\subsection{\label{sec:leve21}Device structure }
The structure investigated in this work is schematized in Fig \ref{Fig_one}. It consists in a graphene ribbon to the sides of which BN flakes are periodically attached. It is thus an alternating arrangement of graphene and BN/graphene/BN (BN/G/BN) sections. Such BN/G/BN ribbons are known to open or broaden the bandgap of free GNRs,\cite{Seol2011d,Tang2012a} and are expected to modify strongly the phonon dispersion, with thus a strong impact on the phonon conductance. In contrast, the electronic conductance in the central graphene ribbon is expected to be weakly affected by the presence of BN flakes. We have focused our study on armchair structures that usually provide higher $ZT$ than zigzag edge ribbons.\cite{Yang2012b} 

\begin{figure*}[hbtp]
\centering
\includegraphics[width=15cm, height=6cm]{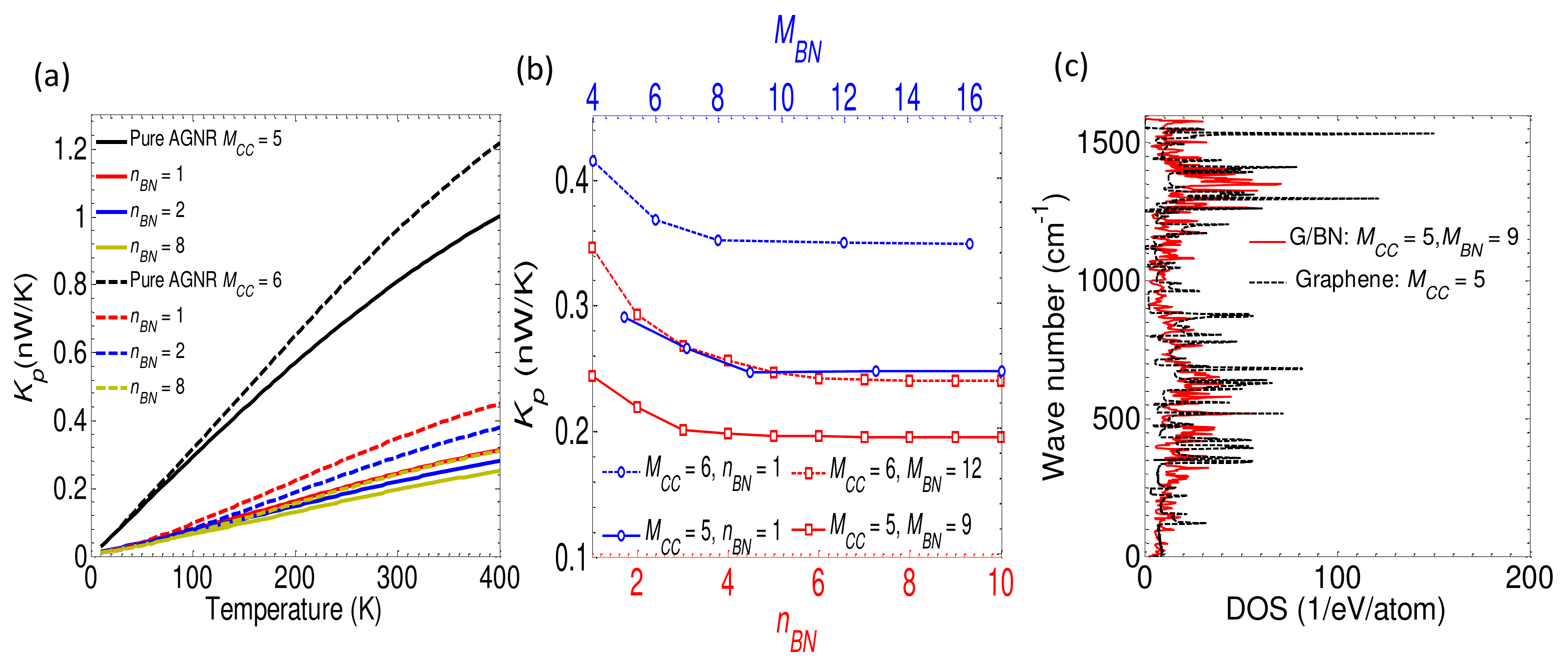}%
\caption{(a) Phonon conductance as a function of temperature for $M_{CC}$ = 5 (solid lines) and $M_{CC}$ = 6 (dashed lines), for different values of the number nBN of BN/G/BN sections (b) Phonon conductance at room temperature for $M_{CC}$ = 5 and $M_{CC}$ = 6 as a function of $n_{BN}$ (red squares) and $M_{BN}$. (blue circles). (c) Phonon density of states in AGNR with $M_{CC}$ = 5 and in BN/G/BN ribbon with $M_{CC}$ = 5 and $M_{BN}$ = 9.}
\label{Fig_two}
\end{figure*}

The width of each sub-region is characterized by the numbers of dimer lines $M_{CC}$ and $M_{BN}$. Along the transport direction, the structure is characterized by the number $n_{BN}$ of BN/G/BN sections and the numbers of unit cells $N_{vc}$ and $N_{BN}$ in graphene and BN/G/BN sections, respectively. For instance in case of the structure sketched in Fig. \ref{Fig_one}, we have $M_{CC} = M_{BN} = 4, N_{vc} = N_{BN} =2, n_{BN} = 2$. The total number of unit cells in the active region is ${N_A} = \left( {{n_{BN}} + 1} \right){N_{vc}} + {n_{BN}}{N_{BN}}$. This active region is connected to BN/G/BN leads having a large bandgap. 
It is assumed that the system can be extended periodically along the $y$ direction, which requires $M_{CC} + M_{BN}$  to be an even number.\cite{Seol2011d}

\subsection{\label{sec:leve22}Methodologies } 
The electronic properties were investigated using a nearest-neighbor tight binding (TB) approach. The corresponding atomistic Hamiltonian takes the general form\cite{Zheng2007a}
\begin{equation}
{H_e} = \sum\limits_i {{\varepsilon _i}\left| i \right\rangle \left\langle i \right|}  - \sum\limits_{\left\langle {i,j} \right\rangle } {{t_{ij}}\left| i \right\rangle \left\langle j \right|} 
\label{eq_two}
\end{equation}
where ${\varepsilon _i}$ is the on-site energy at site $i$ and ${t_{ij}}$ is the hoping energy between atoms at \textit{i}-th and \textit{j}-th sites. Starting from the second-nearest-neighbor TB model of Ref.\cite{Seol2011d}, we made some small changes to work with a simple nearest-neighbor model that reproduces very well the energy band structure of armchair graphene/BN ribbons computed using first principle method.\cite{Seol2011d} The values of the parameters used are listed in table \ref{table_I}.

\begin{table}[b]
\caption{\label{table_I}%
Tight binding parameters for electron study. }
\begin{ruledtabular}
\begin{tabular}{ccdddddd}
$E_{C_A}$&$E_{C_B}$&
\multicolumn{1}{c}{\textrm{$E_{B}$}}&
\multicolumn{1}{c}{\textrm{$E_{N}$}}&
\multicolumn{1}{c}{\textrm{$t_{CC}$}}&
\multicolumn{1}{c}{\textrm{$t_{BN}$}}&
\multicolumn{1}{c}{\textrm{$t_{BC}$}}&
\multicolumn{1}{c}{\textrm{$t_{NC}$}}\\
\hline
eV&eV&\mbox{eV}&\mbox{eV}&\mbox{eV}&\mbox{eV}&\mbox{eV}&\mbox{eV}\\
0.015&-0.015& 1.95 & -1.95 & 2.5 & 2.9 & 2.0 & 2.0 \\
\end{tabular}
\end{ruledtabular}
\end{table}
The phonon properties were computed using the fourth nearest-neighbor force constant (FC) model.\cite{saito1998physical} The general form of the Hamiltonian for the vibrations of atoms was given in\cite{Yamamoto2006} and can be rewritten as 
\begin{equation}
{H_p} = \sum\limits_i {\frac{1}{2}{M_i}\,{{\dot u}_i}^\dag \,{{\dot u}_i}}  + \sum\limits_{\left\langle {i,j} \right\rangle } {\frac{1}{4}\,{{\left( {{u_i} - {u_j}} \right)}^\dag }{K_{ij}}\,\left( {{u_i} - {u_j}} \right)}
\label{eq_three}
\end{equation}

where $M_i$ and $u_i$ are the mass and the time-dependent displacement of the \textit{i}-th atom, respectively, and ${K_{ij}}$ is the $3\times3$ coupling tensor between the \textit{i}-th and \textit{j}-th atoms. Accordingly, in the small displacement limit, the motion equation of the \textit{i}-th atom writes\cite{Mazzamuto2011b,saito1998physical}

\begin{equation}
{M_i}\frac{{{d^2}{u_i}}}{{d{t^2}}} = \sum\limits_{j \ne i} {{K_{ij}}\left( {{u_j} - {u_i}} \right)} 
\label{eq_four}
\end{equation}

For a finite number of atoms, the system of equations \ref{eq_four} may be written in a matrix form. After Fourier transform, it can be rewritten as a function of the angular frequency $\omega$ of the atomic vibrations as ${\omega ^2}U = D\,U$, where $U$ is the column vector formed by the relative displacement vectors ${u_i}$, and $D$ is the dynamic matrix defined as
\begin{equation}
D = \left[ {D_{3 \times 3}^{ij}} \right] = \left[ {\left\{ \begin{array}{l}
 - \frac{{{K_{ij}}}}{{\sqrt {{M_i}{M_j}} }} \;\;\; {\rm{    for }}\;\;\; j \ne i\\
\sum\limits_{n \ne i} {\frac{{{K_{in}}}}{{{M_i}}}}  {\rm{       }}\;\;\;{\rm{for }}\;\;\; j = i
\end{array} \right.} \right]
\label{eq_five}
\end{equation}
It should be noted that the term $\sum\limits_{n \ne i} {\frac{{{K_{in}}}}{{{M_i}}}}$ is a $3\times3$ matrix that plays the same role as the on-site energy   in the case of electrons. However, this term is varying for each atom, depending on the number and nature of couplings with other atoms. It is in contrast with the case of electrons wherein the on-site energy depends only on the atom on this site. The coupling tensor ${K_{ij}}$  between the \textit{i}-th and \textit{j}-th atoms is defined by using a unitary rotation in the plane of the mono-layer graphene/BN structure
\begin{equation}
{K_{ij}} = {U^{ - 1}}\left( {{\theta _{ij}}} \right){K^0}_{ij}U\left( {{\theta _{ij}}} \right)
\label{eq_six}
\end{equation}

where $U\left( {{\theta _{ij}}} \right)$ is the rotation matrix\cite{saito1998physical} and ${\theta _{ij}}$ is the anticlockwise rotating angle formed between the positive direction of the \textit{x}-axis and the vector joining the \textit{i}-th to the \textit{j}-th atoms. Finally, ${K^0}_{ij}$ is the force constant tensor given by
\begin{equation}
{K^0}_{ij} = \left( {\begin{array}{*{20}{c}}
{{\Phi _r}}&0&0\\
0&{{\Phi _{{t_i}}}}&0\\
0&0&{{\Phi _{{t_o}}}}
\end{array}} \right)
\label{eq_seven}
\end{equation}

where ${\Phi _r},{\Phi _{{t_i}}}$ and ${\Phi _{{t_o}}}$ are the force constant coupling parameters in the radial, in-plane and out-of-plane directions, respectively, depending on the level of neighboring. Within a fourth nearest-neighbor, we thus need twelve parameters for graphene and fifteen parameters for BN. In this work the parameters were taken from Ref.\cite{Wirtz2004a} for graphene and from Ref.\cite{Xiao2004} for BN. For the coupling of C atoms with B and N atoms we have taken the average value of the force constant parameters between graphene and h-BN.\cite{Yang2012b} The masses of C, B, N atoms were taken to be equal to $1.994\times10^{-26}, 1.795\times10^{-26}$ and $2.325\times10^{-26}$ kg, respectively.

\begin{figure*}[hbtp]
\centering
\includegraphics[width=13cm, height=12cm]{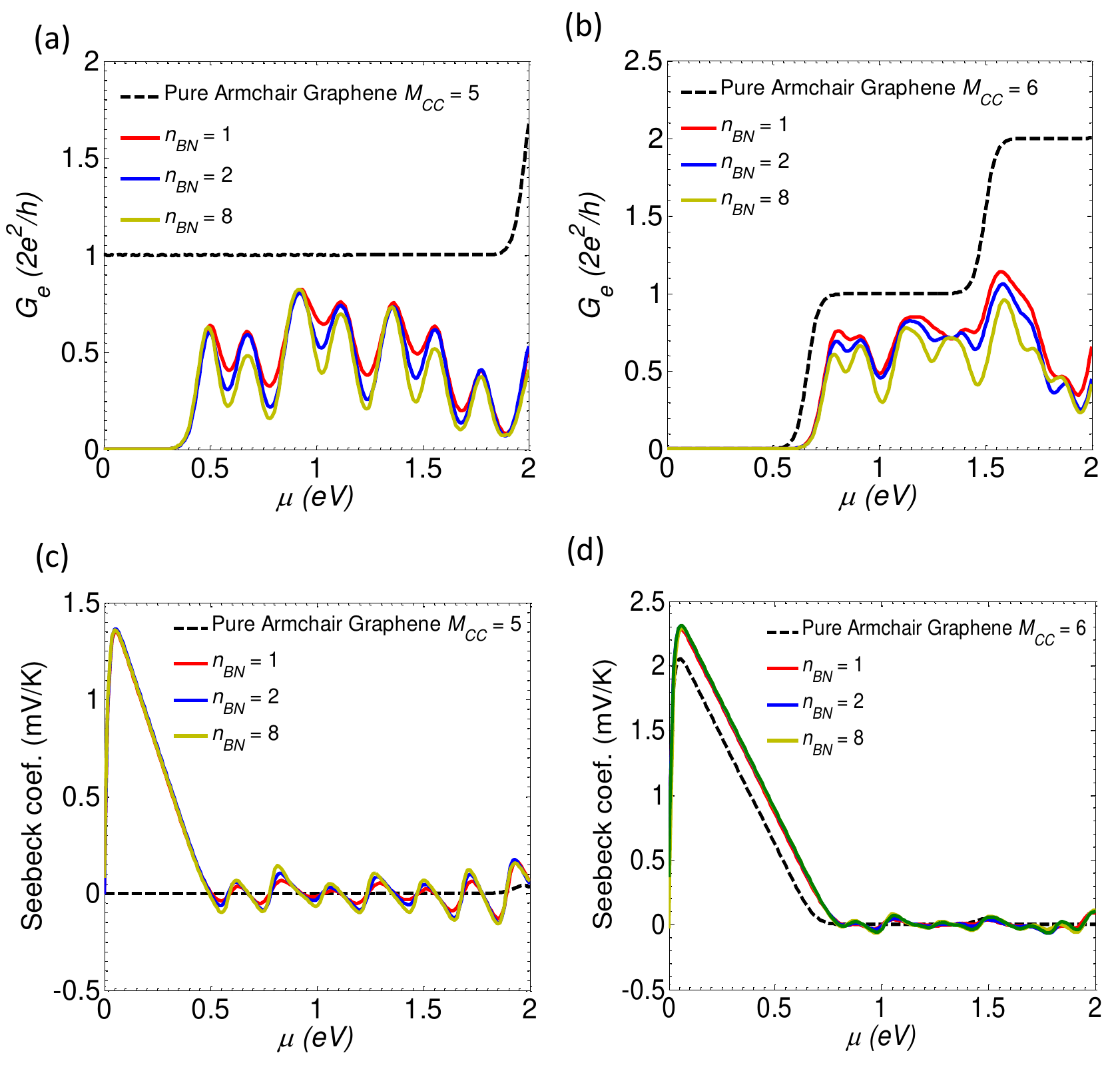}
\caption{(a) and (b) Electrical conductance and (c) (d) Seebeck coefficient for the structures with (a) (c) $M_{CC}$ = 5 and (b) (d) $M_{CC}$ = 6 for different numbers nBN of BN/G/BN sections. Other parameters: $T$ = 300K.}
\label{Fig_three}
\end{figure*}

Due to the similarity between the TB Schrödinger equation for electrons and the FC dynamic equation for phonons, they can formally be solved using the same approach. The Green's function method has been used in the ballistic approximation.\cite{Zhu2014} In the case of electrons the total Hamiltonian can be separated in the device Hamiltonian $H_D$ of the active region, the Hamiltonian $H_{L(R)}$ of the left (right) lead and the coupling term ${\tau _{L(R)}}$ between the device and the left (rigt) lead. First, the surface Green's functions $G_{L(R)}^0$ of the uncoupled leads are calculated using the Sancho's iterative scheme.\cite{Sancho1984} They are used to calculate the self-energies $\Sigma _{L(R)}^s = {\tau _{L(R)}}G_{L(R)}^0\tau _{L(R)}^\dag $ describing the device-lead couplings that take place into the device Green's function $G = {\left[ {E + i\,\eta  - {H_D} - \Sigma _L^s - \Sigma _R^s} \right]^{ - 1}}$ . In the case of phonons a similar expression of the Green's function is found by replacing the energy $E$ of electrons with the square of the phonon angular frequency ${\omega ^2}$. A recursive technique is applied to reduce the size of the device Green's function to the size of the Green's function $G_{11}$ of the first layer.\cite{Lewenkopf2013,Li2008} Then, the electron and phonon transmissions, ${T_e}\left( E \right)$  and ${T_p}\left( \omega  \right)$, can be computed as
\begin{equation}
T = {\rm{Tr}}\left\{ {\Gamma _L^s\left[ {i\left( {{G_{11}} - {G_{11}}^\dag } \right) - {G_{11}}\Gamma _L^s{G_{11}}^\dag } \right]} \right\}
\label{eq_eight}
\end{equation}
where $\Gamma _{L(R)}^s = i\left( {\Sigma _{L(R)}^s - \Sigma {{_{L(R)}^s}^\dag }} \right)$ is the surface injection rate at the left (right) lead. Finally, once the transmission ${T_e}\left( E \right)$ and ${T_p}\left( \omega  \right)$ are calculated, the electrical conductance, the Seebeck coefficient, and the electron and phonon contributions to the thermal conductance can be computed as\cite{0953-8984-27-13-133204}

\begin{equation}
\left\{ \begin{array}{l}
{G_e}\left( {\mu ,T} \right) = {e^2}{L_0}\left( {\mu ,T} \right)\\
S\left( {\mu ,T} \right) = \frac{1}{{e\,T}}\;\frac{{{L_1}\left( {\mu ,T} \right)}}{{{L_0}\left( {\mu ,T} \right)}}\\
{K_e}\left( {\mu ,T} \right) = \frac{1}{T}\;\left[ {{L_2}\left( {\mu ,T} \right) - \frac{{{L_1}{{\left( {\mu ,T} \right)}^2}}}{{{L_0}\left( {\mu ,T} \right)}}} \right]\\
{K_p} = \int\limits_0^\infty  {\frac{{d\omega }}{{2\pi }}\;\hbar \omega \;{T_p}\left( \omega  \right)\;\frac{{\partial n\left( {\omega ,T} \right)}}{{\partial T}}} 
\end{array} \right.
\label{eq_nine}
\end{equation}
where the intermediate function ${L_n}\left( {\mu ,T} \right)$ is defined as
\begin{equation}
{L_n}\left( {\mu ,T} \right) = \frac{2}{h}\int\limits_{ - \infty }^{ + \infty } {dE\;{T_e}\left( E \right)\;{{\left( {E - \mu } \right)}^n}\;\frac{{ - \partial {\kern 1pt} {f_e}\left( {E,\mu ,T} \right)}}{{\partial E}}} 
\label{eq_ten}
\end{equation}
In these equations, ${f_e}\left( {E,\mu ,T} \right)$ is the Fermi distribution function, $n\left( {\omega ,T} \right)$ is the Bose-Einstein distribution function, $T$ is the temperature and µ is the electron chemical potential. More practical forms of the intermediate function and the phonon conductance can be derived as
\begin{equation}
\left\{ \begin{array}{l}
{L_n}\left( {\mu ,T} \right) = \frac{1}{h}\int\limits_{ - \infty }^{ + \infty } {dE\;{T_e}\left( E \right)\;{{\left( {2{k_b}T} \right)}^{n - 1}}\;{g^e}_n\left( {E,\mu ,T} \right)} \\
{K_p} = \frac{{{k_b}}}{{2\pi }}\int\limits_0^\infty  {d\omega \;{T_p}\left( \omega  \right)\;{g^p}\left( {\omega ,T} \right)} 
\end{array} \right.
\label{eq_eleven}
\end{equation}
where ${g^e}_n\left( {E,\mu ,T} \right) = {{{{\left( {\frac{{E - \mu }}{{2{k_b}T}}} \right)}^n}} \mathord{\left/
 {\vphantom {{{{\left( {\frac{{E - \mu }}{{2{k_b}T}}} \right)}^n}} {{{\cosh }^2}\left( {\frac{{E - \mu }}{{2{k_b}T}}} \right)}}} \right.
 \kern-\nulldelimiterspace} {{{\cosh }^2}\left( {\frac{{E - \mu }}{{2{k_b}T}}} \right)}}$, ${g^p}\left( {\omega ,T} \right) = {{{{\left( {\frac{{\hbar \omega }}{{2{k_b}T}}} \right)}^2}} \mathord{\left/
 {\vphantom {{{{\left( {\frac{{\hbar \omega }}{{2{k_b}T}}} \right)}^2}} {{{\sinh }^2}\left( {\frac{{\hbar \omega }}{{2{k_b}T}}} \right)}}} \right.
 \kern-\nulldelimiterspace} {{{\sinh }^2}\left( {\frac{{\hbar \omega }}{{2{k_b}T}}} \right)}}$ and $k_b$ is the Boltzmann constant. Once $G_e$, $S$, $K_e$ and $K_p$ are obtained from equations \ref{eq_nine}, the figure of merit $ZT$ can be computed from equation \ref{eq_one}
 
\section{\label{sec:level3}Results and discussion }
\subsection{\label{sec:level31}Phonon scattering: a key to enhance $ZT$} 
In the following analysis of the thermoelectric performance of the device schematized in Fig. \ref{Fig_one}, unless otherwise stated we will focus mainly on two cases of graphene ribbon width, i.e. $M_{CC}$ = 5 and $M_{CC}$ = 6, that correspond to a metallic and a semiconducting ribbon, respectively, in the case of pure AGNR. Similarly, the BN width $M_{BN}$ will be chosen equal to 9 and 12 for $M_{CC}$ = 5 and 6, respectively, that correspond to the width beyond which the transport properties are unchanged, as will be discussed later. The length of active region can be adjusted by changing the number $n_{BN}$ of BN/G/BN sections, which will be used as a parameter in what follows. Indeed, by increasing this number one may expect to reduce the phonon conductance as a consequence of enhanced phonon scattering. 

\begin{figure}[hbtp]
\centering
\includegraphics[width=9cm, height=9cm]{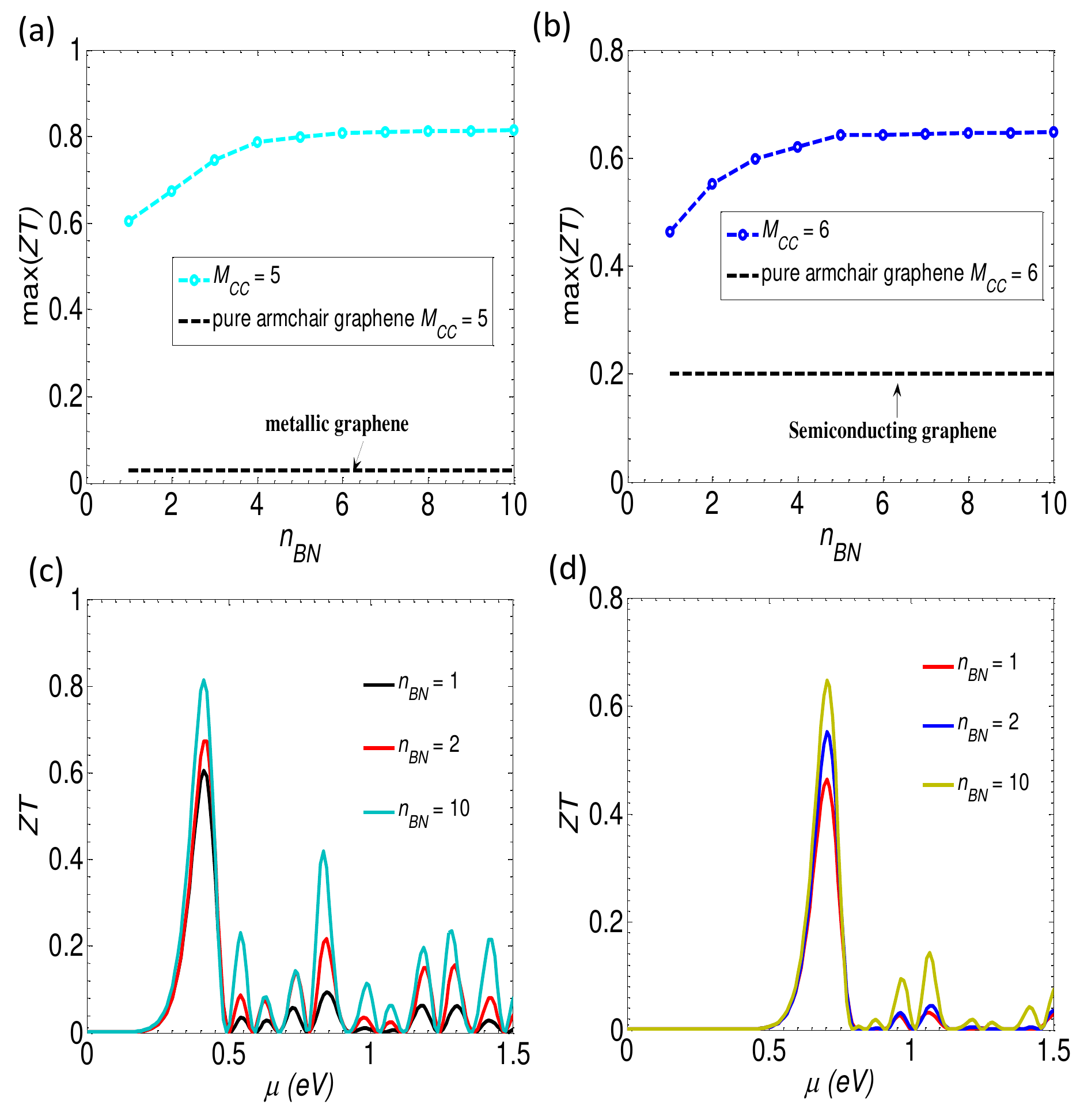}
\caption{(a)(b) maximum of thermoelectric figure of merit as a function of $n_{BN}$ for (a) $M_{CC}$ = 5 and (b) $M_{CC}$ = 6. Thermoelectric figure of merit $ZT$ as a function of chemical potential µ for (c) $M_{CC}$ = 5 and (d) $M_{CC}$ = 6. $T$ = 300K.}
\label{Fig_four}
\end{figure}

In Fig. \ref{Fig_two}(a) we plot the phonon conductance for $M_{CC}$ = 5 and $M_{CC}$ = 6 as a function of the temperature, for different numbers $n_{BN}$ of BN/G/BN sections in the active device. We also plot in black lines for comparison the phonon conductance of the corresponding pristine AGNRs. For both widths we obtain a strong reduction of $K_p$ in hybrid structures compared to graphene. Additionally, it can be seen that the phonon conductance decreases further when $n_{BN}$ increases. For instance, in the case of $M_{CC}$ = 5 at room temperature (Fig. \ref{Fig_two}(b), red squares; solid line) i.e. it drops from 0.244 nW/K for $n_{BN}$ = 1 to 0.196 nW/K for $n_{BN}$ = 5. Then the phonon conductance saturates when the active region is long enough because the system tends to behave as an infinite superlattice system. Finally, for $n_{BN}$ = 10 at room temperature, $K_p$ is about 75\% smaller than in the corresponding pristine graphene ribbon. Similarly, for $M_{CC}$ = 6 the phonon conductance falls from 0.96 nW/K (pristine AGNR) to 0.346 nW/K ($n_{BN}$ = 1) and even 0.24 nW/K ($n_{BN}$ = 10). A similar behavior is observed when the width $M_{BN}$ (blue circles in Fig. \ref{Fig_two}(b) of the BN flakes increases. A minimum value of about 10 is required to reach the minimum of conductance.

This dramatic reduction of phonon conductance is not only due to phonon scattering at graphene/BN boundaries,\cite{KInacI2012} but above all to the scattering induced by the mismatch of modes between pristine graphene and BN/G/BN sections that reduces strongly the phonon transmission. This mismatch of modes is illustrated in Fig. \ref{Fig_two} (c) where we plot the phonon density of states calculated in both sections assumed to be infinite.

The electrical conductance $G_e$ and Seebeck coefficient $S$ are plotted in Fig. \ref{Fig_three} as a function of chemical energy µ in both cases of $M_{CC}$ = 5 and 6 and, for comparison, in the case of pristine GNR (black dashed lines). In Figs. \ref{Fig_three}(a) and \ref{Fig_three}(c), i.e. for $M_{CC}$ = 5 corresponding to a metallic AGNR, we can observe the opening of a significant conduction energy gap, which manifests itself not only in the vanishing of $G_e$ at low energy, but also in the strong enhancement of $S$, compared to its pristine GNR counterpart, whatever the value of $n_{BN}$. In the case of $M_{CC}$ = 6, for which the pristine GNR is semiconducting, the bandgap is slightly broadened, which yields a small enhancement of the Seebeck coefficient, independently of the number of periods $n_{BN}$ (Figs. \ref{Fig_three}(b) and \ref{Fig_three}(d)). It is remarkable that in contrast with the phonon conductance, the electronic conductance is not severely degraded with respect to the pristine GNR, whatever the value of $n_{BN}$, at least at low energy (first step of transmission) at which most of the conduction takes place. Regarding thermoelectric performance, it is definitely a strong advantage of this type of nanostructuring design. It should be noted also that the maximum Seebeck coefficient is higher in the case $M_{CC}$ = 6, corresponding to a larger bandgap. This behavior is in agreement with the general result that for any material the maximum value of $S$ should be proportional to the bandgap,\cite{Goldsmid1999} which has been observed in several other cases.\cite{Mazzamuto2011b,Nguyen2015} 

As a consequence of the strong reduction of phonon conductance, together with high power factor, good thermoelectric figures of merit are expected in theses hybrid structures. Since the phonon conductance appears to be strongly dependent on the alternating arrangement of graphene and BN/G/BN sections in the active region, the results should depend also on the lengths $N_{vc}$ and $N_{BN}$ of these sections. Our investigations led us to the conclusion that the maximum value of $ZT$ is reached when ${{{N_{vc}}} \mathord{\left/ {\vphantom {{{N_{vc}}} {{N_{BN}} \approx 1}}} \right. \kern-\nulldelimiterspace} {{N_{BN}} \approx 1}}$, which is in agreement with the conclusion of Yang et al.\cite{Yang2012b}  in the case of a transverse superlattice of graphene and BN sections. We found also that for $M_{CC}$ = 5 and $M_{CC}$ = 6 the maximum of $ZT$ is obtained for $N_{vc}$ = $N_{BN}$ = 8. However, for $M_{CC}$ = 3 the best $ZT$ is achieved for $N_{vc}$ = 6 and $N_{BN}$ = 4. 

In both cases of $M_{CC}$ = 5 and $M_{CC}$ = 6, we plot in Fig. \ref{Fig_four}  the maximum value of $ZT$ as a function of the number $n_{BN}$ of BN/G/BN sections in the active region (Figs. \ref{Fig_four}(a) and \ref{Fig_four}(b)), and the evolution of $ZT$ as a function of the chemical potential in the leads, for different values of $n_{BN}$ (Figs. \ref{Fig_four}(c) and \ref{Fig_four}(d). In all cases, the case of pristine GNR of same width is shown for comparison. Due to the metallic behavior of AGNRs with $M_{CC}=3p+2$, the AGNR with $M_{CC}$ = 5 exhibits a very small Seebeck coefficient (see Fig. \ref{Fig_three}(c)) and a $ZT$ of about 0.03. Thanks to the bandgap opening and the dramatic reduction of phonon conductance, the figure of merit is strongly enhanced in the hybrid structure (see Figs. \ref{Fig_four}(a) and \ref{Fig_four}(c)), reaching 0.6 for $n_{BN}$ = 1 and 0.81 for longer devices with $n_{BN}> 5$ . More interestingly, it should be remarked that the peak values of $ZT$ are observed here for the quite low chemical energy µ = 0.41 eV, which is much smaller and more accessible than in the case of the transverse graphene/BN superlattice.\cite{Yang2012b} 

\begin{figure}[hbtp]
\centering
\includegraphics[width=7.5cm, height=5.5cm]{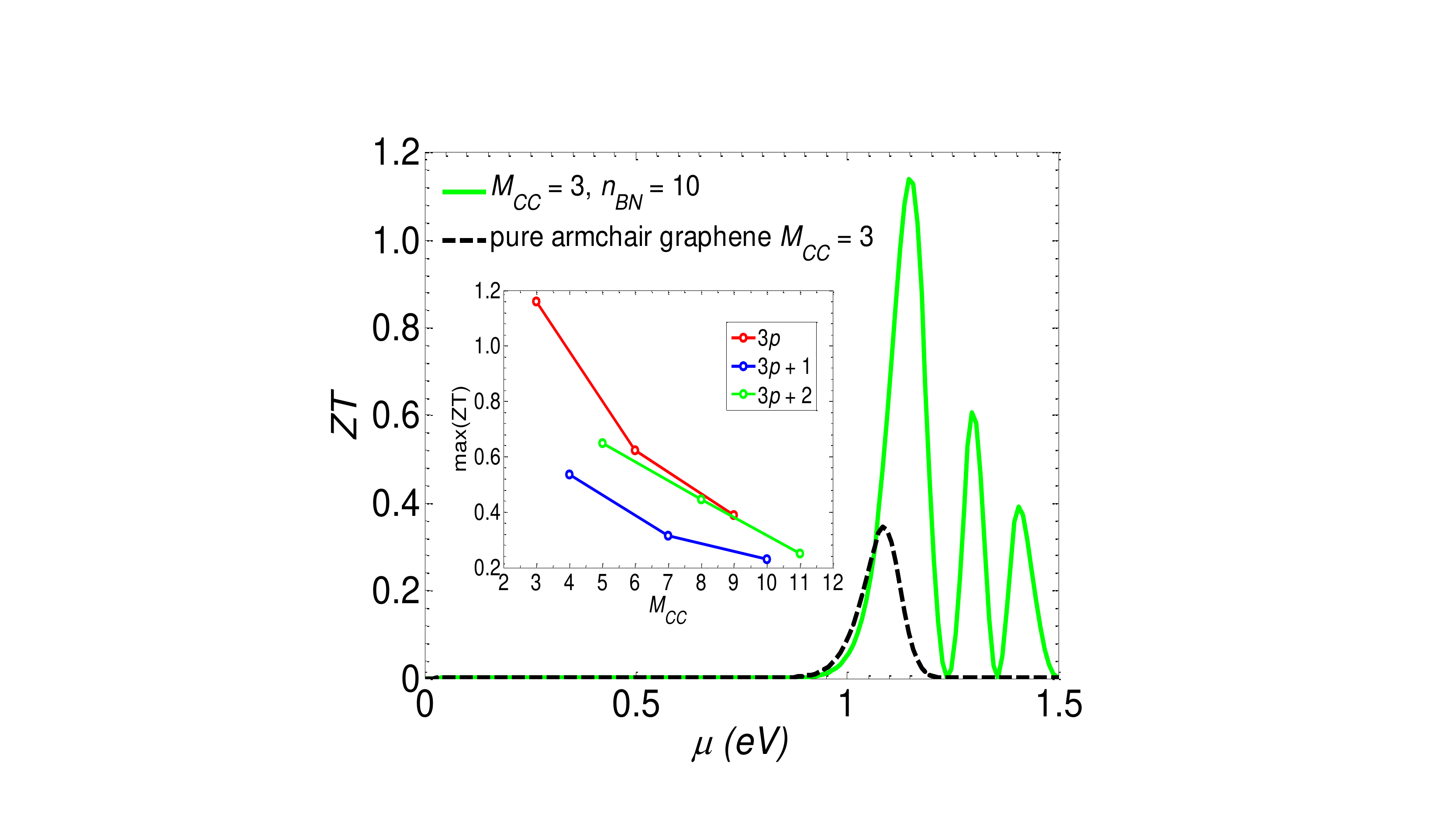}
\caption{$ZT$ as a function of chemical potential for the thinnest ribbon $M_{CC}$ = 3 in the cases of pure GNR (dashed line) and hybrid heterostructure with $M_{BN}$ = 11 (solid line), $M_{vc}$ = 6, $N_{BN}$ = 4, $n_{BN}$ = 10, $T$ = 300 K. Inset: Dependence of $ZT_{max}$ on the GNR width $M_{CC}$ depending on whether $M_{CC}$ can be written in the form $3p, 3p+1$ or $3p+2$, with $p$ is an integer number. For the inset, $M_{BN}$ = 10, 11 for even, odd $M_{CC}$, respectively and $M_{vc} = N_{BN}$ = 5, $n_{BN}$ = 10.}
\label{Fig_five}
\end{figure}

For $M_{CC}$ = 6, corresponding to a semiconducting AGNR, the peak value of $ZT$ reaches 0.65, which is about 3.25 times greater than in the AGNR counterpart (Fig. \ref{Fig_four}b). Due to the larger bandgap, this peak value is obtained for a chemical potential µ = 0.7 eV, i.e. higher than in the case $M_{CC}$ = 5.  
In the case of the smallest graphene ribbon $M_{CC}$ = 3 with a large conduction gap, a very high $ZT$ can be achieved, as shown in Fig \ref{Fig_five}. The maximum value $ZT_{max} = 1.14$ is reached for $n_{BN}$ = 10 and the chemical potential µ = 1.15 eV. It is again an improvement by a factor of 3.25 compared to the value of 0.35 in the case of pristine AGNR of same width (black dashed line). 

The inset in Fig. \ref{Fig_five} shows the evolution of $ZT$ as a function of the width $M_{CC}$ of the main GNR channel. In principle, the maximum of $ZT$ reduces when increasing $M_{CC}$ due to the increase of phonon conductance. However, the behavior is not monotonic and the value of $ZT$ is smaller for the group of structures with $M_{CC} = 3p+1$ compared to the groups with $M_{CC} = 3p$ and $M_{CC} = 3p+2$. 

\begin{figure}[hbtp]
\centering
\includegraphics[width=7cm, height=5.5cm]{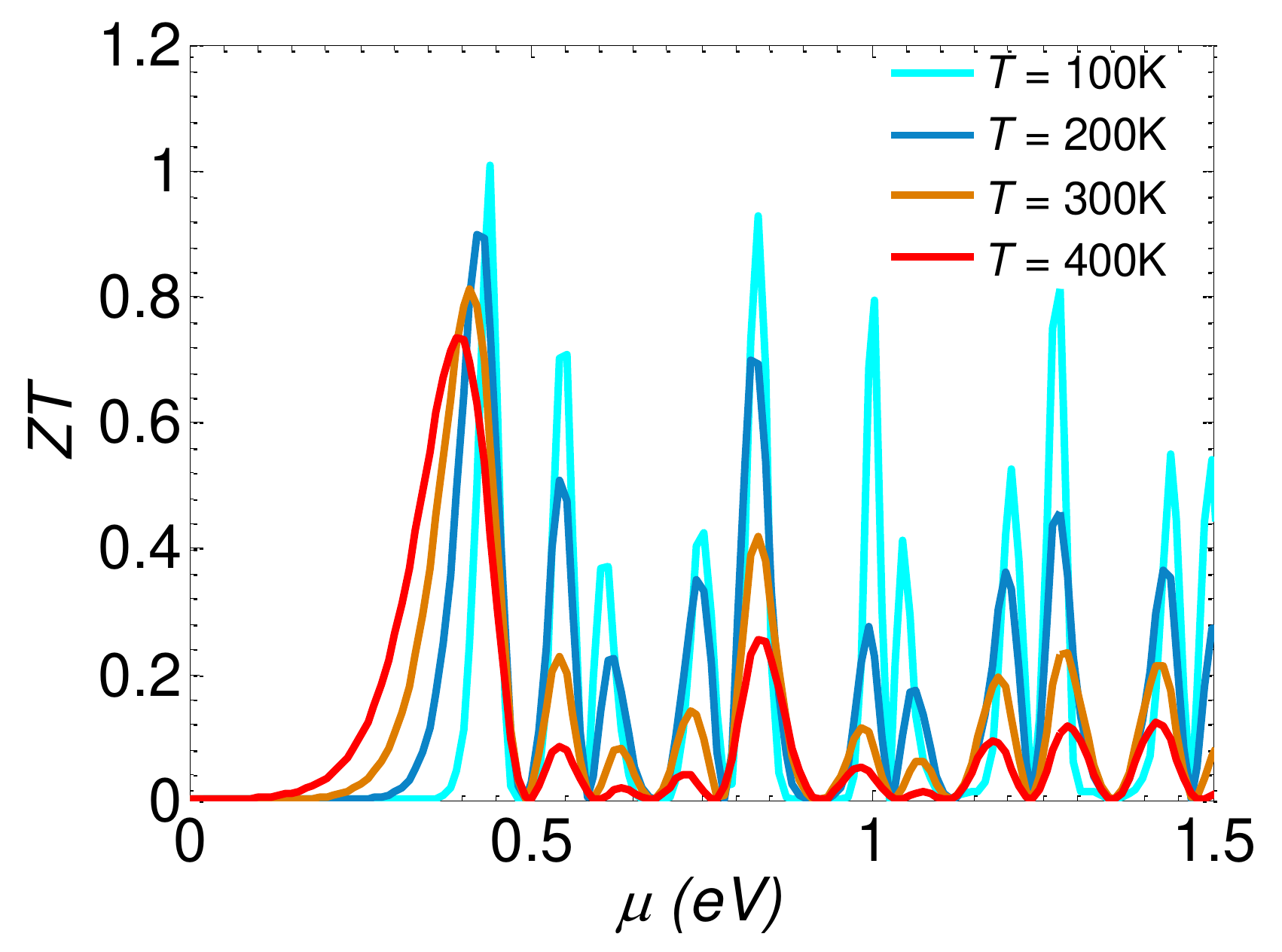}
\caption{$ZT$ as a function of chemical potential for different temperatures ranging from 100K to 400K for the heterostructure with $M_{CC}$ = 5. Here $M_{BN}$ = 9, and $n_{BN}$ = 10.}
\label{Fig_six}
\end{figure}

\begin{figure*}[hbtp]
\centering
\includegraphics[width=15cm, height=12cm]{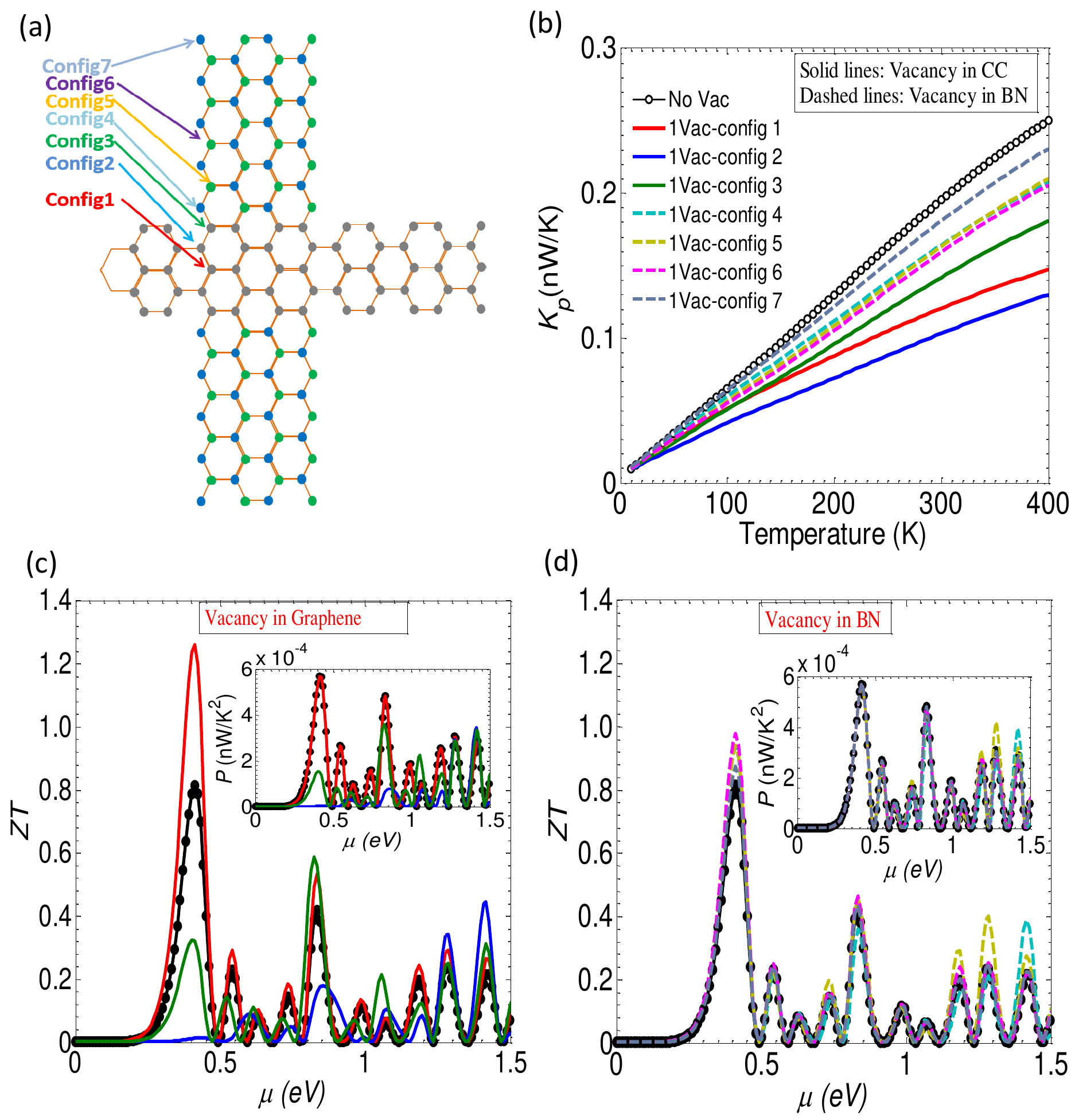}
\caption{Effect of a single vacancy on the thermoelectric properties for the structure with $M_{CC}$ = 5. (a) The seven vacancy configurations investigated here. (b) Phonon conductance for the different vacancy configurations. $ZT$ and (inset) thermal power for the different configurations of vacancy in (c) graphene and (d) BN. Symbols correspond to results without any vacancy. $T$ = 300 K.}
\label{Fig_seven}
\end{figure*}

In Fig. \ref{Fig_six} we plot $ZT$ as a function of µ for different temperatures ranging from 100 K to 400 K in the case of the structure with $M_{CC}$ = 5. Starting from room temperature (orange curve) with $ZT_{max} = 0.81$, when the temperature decreases to 200 K and 100K, the peak value of $ZT$ increases to $ZT_{max} = 0.9$ and $ZT_{max} = 1.01$ , respectively. When increasing the temperature to 400 K, the figure of merit reduces to $ZT_{max} = 0.74$. This reduction of $ZT_{max}$ when the temperature increases is accompanied by a shift of the peak to the left, i.e. to smaller chemical potentials. It is due to the broadening of the functions ${g^e}_n\left( {E,\mu ,T} \right)$  at increasing temperature, which moves the highest peak of Seebeck coefficient to lower values of µ.

\subsection{\label{sec:level32}Effect of vacancies} 
In this next sections we will discuss the effects of vacancies on the thermoelectric performance of the hybrid structure. We will focus our investigations on the structure with $M_{CC}$ = 5 for which we observe high $ZT$ at relative low chemical potential. 

It has been shown that edge disorder and inner vacancies may have a significant influence on the thermoelectric performance of GNRs.\cite{Sevincli2010,Mazzamuto2012}  In particular, depending on the vacancy position the figure of merit $ZT$ can be larger than in a perfect GNR. Most of works on vacancy effects in graphene were focused on either random or periodic arrangement of vacancies. To understand better the influence of vacancies in our hybrid structure, we will investigate the effect of the position of a vacancy. We assume that it may be possible to control experimentally this position by focused electron beam technique.\cite{Rodriguez-Manzo2009,Robertson2012}  We will first consider different configurations of a single vacancy from the middle of channel to the edge, and then the effect of a few vacancies. In our structure of Fig. \ref{Fig_one}, the position of a vacancy is determined by three indices, i.e. (i) $n_{vac}$ that is the position of the unit cell to which belongs the vacancy, running from 1 to $N_A$, (ii) $L_{id-vac}$ that is the layer index of a layer in the cell, equal to 1 or 2, and (iii) $m_{vac}$ that is the position of the vacancy in the layer, ranging from 1 to $M$, where $M = M_{CC} + 2M_{BN}$ is the total number of dimer lines along the width. For the structure with $M_{CC} = 5, M_{BN} = 9, N_{vc} = N_{BN} = 8, n_{BN} = 10$, we will consider specifically seven different configurations of a single vacancy at $m_{vac} = 12, 13, 14$ (vacancy in graphene) and 15, 16, 18, 23 (vacancy in BN) in layer $L_{id-vac}$ = 1 of the cell $n_{vac}= 25$ which is the first cell of the second BN/G/BN section with respect to the left contact. These positions are sketched in Fig. \ref{Fig_seven}(a). 

\begin{figure}[hbtp]
\centering
\includegraphics[width=7cm, height=5cm]{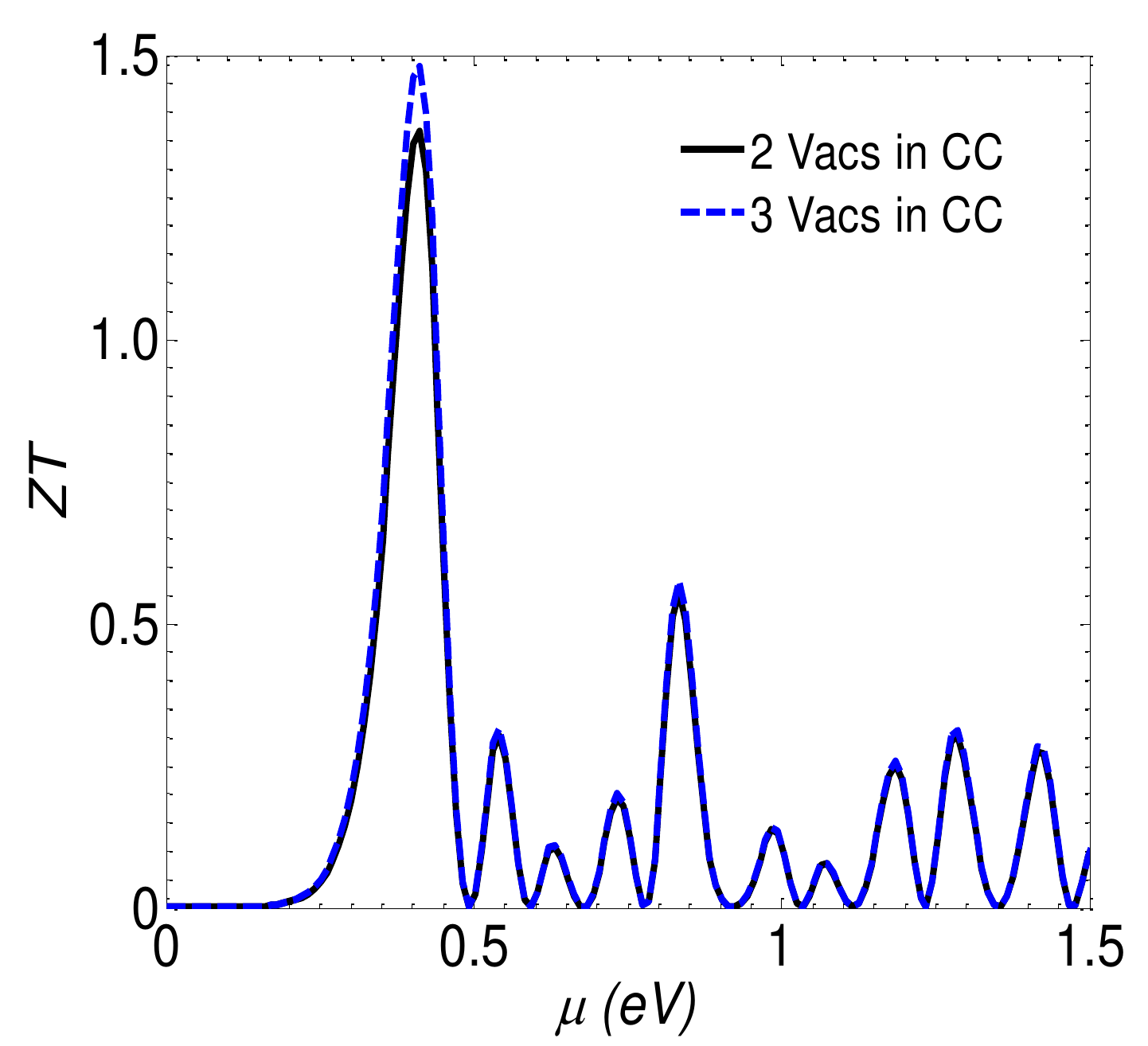}
\caption{Effect of two and three vacancies in the middle of the graphene (CC) ribbon on the figure of merit. $M_{CC}$ = 5, $T$ = 300 K.}
\label{Fig_eight}
\end{figure}

The phonon conductance is plotted in Fig. \ref{Fig_seven}(b) for the different positions of the vacancy in graphene (solid lines) or in BN region (dashed lines) and compared with the case of no vacancy (circles). We observe a strong fall of conductance when the vacancy is located in the center of the graphene ribbon (red line, config-1). It is even stronger when the vacancy is on the next lattice site (blue line, config-2), with a reduction by almost a factor of 2 compared to the no-vacancy configuration. When the vacancy is at the edge of the graphene ribbon, the reduction of phonon conductance is smaller (green line, config-3). When the vacancy is located in the BN region, we still observe a reduction of phonon conductance, but more limited than when it is in graphene and less sensitive to the position of the vacancy. 

In Fig. \ref{Fig_seven}(c) and \ref{Fig_seven}(d) we plot the thermoelectric figure of merit $ZT$ as a function of µ for the different configurations of the vacancy in graphene and in BN, respectively. The corresponding power factor $P$ is displayed in inset. In comparison with the case without vacancy, we observe a remarkable enhancement of $ZT$ up to 1.26 if the vacancy is in configuration 1, while we see a degradation of $ZT$ in the other two cases of vacancy in graphene. Indeed, in spite of the strong reduction of phonon conductance, the configuration 2 surprisingly leads to smaller $ZT$, which is due to a very small power factor, as shown in inset of Fig. \ref{Fig_seven}(c). In contrast, the power factor is not at all affected when the vacancy is in the center of the graphene ribbon, which is also quite surprising. This behavior can be understood from the analysis of the density in the ribbon that is strongly dependent on the lattice site position, as shown in previous works.\cite{Tran2014a,Wilhelm2014}  By analyzing the current density in AGNRs Whilhelm et al. shown that as a consequence of quantum confinement in the transverse flow direction, streamlines appear with a threefold periodicity in transverse site position, separated by stripes of almost vanishing flow.\cite{Wilhelm2014} We observe a similar behavior in the graphene part of the structures studied in the present work. Starting from an edge of the GNR, i.e. $m = 1$ at the edge, the density is minimum for the positions $m$ that are multiple of 3. In the particular case of the ribbon of width $M_{CC}$ = 5 studied here, the density in the center of the ribbon ($m = 3$) is several order of magnitude smaller than at other positions $m = 1$ (maximum) and $m = 2$. Hence, if there is a vacancy at this position, i.e. in configuration 1, the effect on the density and the conductance of the ribbon is negligible, which explains that the power factor is the same as in in the no-vacancy case, leading to an enhancement of $ZT$ as a consequence of the reduced phonon conductance. In contrast, if the vacancy is at any other position in the GNR, the density and the conductance are strongly affected, which also degrades $ZT$ in spite of the reduction of phonon conductance.

When the vacancy is in the BN region (Fig. \ref{Fig_seven}(d)) the power factor is almost not influenced at low energy, which results in a small enhancement of $ZT$, as a consequence of the reduction of phonon conductance observed in Fig. \ref{Fig_seven}(b). In this case, we always observe an enhancement of $ZT$ for any vacancy in a BN region. This is an interesting property of G/BN heterostructures compared to pure graphene ribbons in which a vacancy may either degrade or improve $ZT$. 

To fully exploit the influence of vacancies in the center of the graphene ribbon, we have made simulations with two or three vacancies randomly distributed in the center line of the structure with $M_{CC}$ = 5 and $n_{BN}$ = 10. As shown in Fig. \ref{Fig_eight}, $ZT$ can be further enhanced and reaches 1.48 in the case of three vacancies, as a consequence of additional degradation of phonon conductance. 

\section{\label{sec:level4}Conclusions }
In this article we have demonstrated excellent thermoelectric properties of in-plane heterostructures based on armchair GNR with appropriate engineering of graphene/h-BN interfaces. By attaching periodically BN flakes to the GNR, it is indeed possible (i) to reduce strongly the phonon conductance compared to pure GNR and (ii) to open or broaden the conduction gap which enhances the Seebeck coefficient. Additionally, the electron conductance is remarkably weakly affected. It finally results in a large enhancement of the thermoelectric figure of merit $ZT$. In the case of perfect ribbons, the maximum value $ZT$max of 1.14 is reached for the thinnest ribbon with $M_{CC}$ = 3 carbon dimer lines along the width. For larger ribbons with $M_{CC}$ = 5 and $M_{CC}$ = 6 we can obtain $ZT_{max}$  of 0.81 and 0.65, respectively. Interestingly, in the case $M_{CC}$ = 5 the peak of $ZT$ is reached at relatively low chemical potential µ = 0.41 eV. When reducing the temperature from 300 K to 100 K, $ZT_{max}$  increases from 0.81 to 1.01. The investigation has also shown that engineering vacancies in graphene or BN regions may allow to further enhance $ZT$ thanks to additional reduction of phonon conductance. However, the electron conductance is very sensitive to the vacancy position in graphene. The vacancy position must be chosen carefully at a lattice site where the electron density is small to avoid unacceptable degradation of the power factor. In the case $M_{CC}$ = 5, it is shown that by generating some vacancies in the center of the GNR, i.e. where the charge density is negligible in a perfect ribbon, the power factor is not affected and $ZT_{max}$ reaches up to 1.48 at room temperature.

\begin{acknowledgments}
This work was partially supported by the French ANR through project NOE (12JS03-006-01).
\end{acknowledgments}


\nocite{*}

\bibliography{ZTinGBN}

\end{document}